\documentclass[aps,onecolumn]{revtex4}
\usepackage{caption2}
\usepackage{float}
\usepackage{amsmath}
\usepackage{amssymb}
\usepackage{graphicx}

\begin{document}

\title{Multidimensional \textit{semi-gap} solitons in a periodic potential}
\author{Bakhtiyor B. Baizakov${1}$, Boris A. Malomed${2}$ and Mario
Salerno${1}$\\
${1}$ Dipartimento di Fisica "E. R. Caianiello", Universit{\'a} di
Salerno, via S. Allende, I-84081 Baronissi (SA), Italy \\
${2}$ Department of Interdisciplinary Studies, School of
Electrical
Engineering, Faculty of Engineering, \\
Tel Aviv University, Tel Aviv 69978, Israel}

\begin{abstract} The existence, stability and other dynamical
properties of a new type of multi-dimensional (2D or 3D) solitons
supported by a transverse low-dimensional (1D or 2D, respectively)
periodic potential in the nonlinear Schr\"{o}dinger equation with
the self-defocusing cubic nonlinearity are studied. The equation
describes propagation of light in a medium with normal
group-velocity dispersion (GVD). Strictly speaking, solitons
cannot exist in the model, as its spectrum does not support a true
bandgap. Nevertheless, the variational approximation (VA) and
numerical computations reveal stable solutions that seem as
completely localized ones, an explanation to which is given. The
solutions are of the gap-soliton type in the transverse
direction(s), in which the periodic potential acts in combination
with the diffraction and self-defocusing nonlinearity.
Simultaneously, in the longitudinal (temporal) direction these are
ordinary solitons, supported by the balance of the normal GVD and
defocusing nonlinearity. Stability of the solitons is predicted by
the VA, and corroborated by direct simulations.
\end{abstract}

\maketitle

\section{Introduction}

Recently, a variety of two- and three-dimensional (2D and 3D)
solitons have been investigated in models based on the nonlinear
Schr\"{o}dinger (NLS) or Gross-Pitaevskii (GP) equations with a
spatially periodic potential and cubic nonlinearity, see a review
\cite{review}. The physical models of this type emerge in the
context of Bose-Einstein condensation (BEC)
\cite{multi-d,bmsEPL,Yang,Estoril,bms,mih}, where the periodic
potential is created as an optical lattice (OL), i.e.,
interference pattern formed by coherent beams illuminating the
condensate, and in nonlinear optics, where similar models apply to
photonic crystals \cite{PhotCryst}. A different but allied setting
is provided by a cylindrical OL (``Bessel lattice"), which can
also support stable 2D \cite{Kartashov} and 3D \cite{mihalache}
solitons. Additionally, models combining a periodic lattice
potential and saturable nonlinearity give rise to 2D solitons,
that were predicted in Ref. \cite{PhotorefrPrediction} and
observed in several experiments in photorefractive media,
including fundamental solitons \cite{Photorefr2Dsolitons} and
vortices \cite{PhotorefrVortices}. It is also relevant to mention
that experimental observation of spatiotemporal self-focusing of
light in silica waveguide arrays, in the region of anomalous
group-velocity dispersion (GVD), was reported in Ref.
\cite{Cheskis}.

In models with the cubic nonlinearity, these solutions were
investigated in a quasi-analytical form, which combines the
variational approximation (VA) \cite{va} to predict the shape of
the solitons, and the Vakhitov-Kolokolov (VK) criterion \cite{vk}
to examine their stability. Final results were provided by
numerical methods, relying upon direct simulations of the
underlying NLS/GP equations. A conclusion obtained by means of
these methods is that, unlike their 1D counterparts,
multi-dimensional solitons in periodic potentials can exist only
in a limited domain of the $\left( N,\varepsilon \right) $ plane,
where $N$ and $\varepsilon $ are the norm of the solution and
strength of the OL potential, respectively. The most essential
limitation on the existence domain of 2D solitons is that $N$
cannot be too small (in a general form, a minimum value of the
norm, as a necessary condition for the existence of 2D solitons
supported by lattice potentials, was discussed in Ref.
\cite{efremidis}). Unlike it, $\varepsilon $ may be arbitrarily
small, as even at $\varepsilon =0$ the 2D\ NLS equation has a
commonly known weakly unstable solution in the form of the
\textit{Townes soliton}, at a single value of the norm,
$N=N_{\mathrm{T}}$ \cite{Townes} ($N_{\mathrm{T}}\approx 11.7$ for
the NLS equation in the usual 2D form, $iu_{t}+\nabla
^{2}u+|u|^{2}u=0$). Small finite $\varepsilon $ gives rise to a
narrow stability region,
\begin{equation}
0<N_{\mathrm{T}}-N<\left( \Delta N\right) _{\max }\sim \varepsilon
\label{nearTownes}
\end{equation}for the 2D solitons \cite{bmsEPL}. Crossing the lower border of the
existence domain (\ref{nearTownes}) leads to disintegration of the localized
state into linear Bloch waves (radiation) \cite{bs}.

In the case of the attractive cubic nonlinearity (which
corresponds to BEC where atomic collisions are characterized by a
negative scattering length, while this is the case of the normal,
self-focusing Kerr effect), 2D and 3D solitons can be stabilized
not only by the potential lattice whose dimension is equal to that
of the ambient space, but also by \textit{low-dimensional}
periodic potentials, whose dimension is smaller by one, i.e., 2D
and 3D solitons can be stabilized by a quasi-1D \cite{Estoril,bms}
or quasi-2D \cite{Estoril,bms,mih} OL, respectively [in the former
case, the qualitative estimate (\ref{nearTownes}) for the width of
the stability region at small $\varepsilon $ is correct too];
however, 3D solitons cannot be stabilized by a quasi-1D lattice
potential \cite{Estoril,bms} [this is possible if the 1D potential
is applied in combination with the \textit{Feshbach-resonance
management}, i.e., periodic reversal of the sign of the
nonlinearity coefficient \cite{Warsaw}, or in combination with
\textit{dispersion management}, i.e., periodically alternating
sign of the local GVD coefficient \cite{Michal}]. Solitons can
exist in such settings because the attractive nonlinearity
provides for stable self-localization of the wave function in the
free direction (one in which the low-dimensional potential does
not act), essentially the same way as in the 1D NLS equation, and,
simultaneously, the lattice stabilizes the soliton in the other
directions (in the 3D model with the quasi-1D OL potential, the
self-localization in the transverse 2D subspace, where the
potential does not act, is possible too, but the resulting soliton
is unstable, the same way as the above-mentioned Townes soliton).
An important aspect of settings based on the low-dimensional OL
potentials is mobility of the solitons along the free direction,
which opens the way to study collisions between solitons and
related dynamical effects \cite{bms}.

In the case of defocusing nonlinearity, which corresponds to a positive
scattering length in the BEC, or self-defocusing nonlinearity in optics
(negative Kerr effect), the soliton cannot support itself in the free
direction. Localization in that direction may be provided by an additional
external confining potential; however, the resulting pulse is not a true
multidimensional soliton, but rather a combination of a \textit{gap soliton}
(a weakly localized state created by the interplay of the repulsive
nonlinearity and periodic potential \cite{multi-d}),which was recently
created experimentally in a 1D BEC \cite{Oberthaler}) in the direction(s)
affected by the OL, and of a \textit{Thomas-Fermi state}, directly confined
by the external potential in the remaining direction \cite{bms}.

Thus, no soliton can be supported by a low-dimensional lattice in
the BEC model (GP equation) with self-repulsion (the latter
corresponds to the most common situation in the experiment
\cite{Oberthaler}). On the other hand, a new possibility may be
considered in terms of nonlinear optics. Indeed, one may combine
three physically relevant ingredients, viz., (i) an effective
periodic potential in the transverse direction(s), while the
medium is uniform in the propagation direction, (ii)
self-defocusing nonlinearity, and (iii) normal GVD. The latter is
readily available, as most optical materials feature normal GVD,
in compliance with its name. As concerns the negative cubic
nonlinearity, it is possible in semiconductor waveguides, or may
be engineered artificially, through the cascading mechanism, in a
quadratically nonlinear medium with a proper longitudinal
quasi-phase-matching \cite{QPM}. Also quite encouraging for the
study of multidimensional solitons proposed in this work are
recent observations of 1D \cite{jason1} and 2D
\cite{Photorefr2Dsolitons} solitons in optically induced waveguide
arrays (photonic lattices) with \emph{self-defocusing}
nonlinearity.

The setting outlined above can be realized in both 2D and 3D
geometry, where the necessary transverse modulation of the
refractive index is provided, respectively, by the transverse
structure in a planar photonic-crystal waveguide, or in a
photonic-crystal fiber. To the best of our knowledge, in either
case the model is a novel one. A soliton in this medium, if it
exists, will be of a \textit{mixed type}: in the transverse
direction(s), it is, essentially, a 1D or 2D spatial gap soliton,
supported by the combination of the effective periodic potential
and self-defocusing nonlinearity, while in the longitudinal
direction it is a temporal soliton of the ordinary type, which is
easily sustained by the joint action of the self-defocusing
nonlinearity and normal GVD. Thus, one may anticipate stable
spatiotemporal\textit{\ solitons}, alias ``light bullets", in this
model. Due to their mixed character, they may be called
\textit{semi-gap solitons}. The issue is of considerable interest
in view of the lack of success in experiments aimed at the
creation of ``bullets" in more traditional nonlinear-optical
settings \cite{review}. The only earlier proposed scheme for the
stabilization of 2D spatiotemporal optical solitons in periodic
structures, that we are aware of, assumed the use of a planar
waveguide with constant self-focusing nonlinearity and
longitudinal dispersion management \cite{WarsawDM}.

On the other hand, it is necessary to stress that, rigorously
speaking, completely localized solutions cannot exist in the
present model: its linear spectrum cannot give rise to any true
bandgap, in which genuine solitons could be found (see below);
instead, one may expect the existence of quasi-solitons,
consisting of a well-localized ``body" and small nonvanishing
``tails" attached to it. Nevertheless, we will produce families of
solutions which seem as stable perfectly localized objects. This
is possible because the ``tails" may readily turn out to be so
tiny that they remain completely invisible in numerical results
(possibly being smaller than the error of the numerical scheme),
and, of course, they will be invisible in any real experiment. An
explanation to this feature is provided by the fact that bandgaps,
which ``almost exist" in the system's spectrum, do not exist in
the strict sense because they are covered by linear modes with
very large wavenumbers. As shown in Ref. \cite{Isaac}, in this
case the amplitude of the above-mentioned tails (which are
composed of the linear modes with very large wavenumbers) is
exponentially small. In fact, families of \emph{stable}
``practically existing" solitons in a second-harmonic-generating
system with opposite signs of the GVD at the fundamental-frequency
and second harmonics, where solitons cannot exist in the rigorous
mathematical sense, were explicitly found in that system, in both
multi- \cite{Isaac} and one- \cite{Kale} dimensional settings.
Implicitly (without discussion of this issue), ``practically
existing" solitons (although, in this case, they were unstable
against small perturbations) were also found in a recent work
\cite{Alejandro}, which was dealing with a 2D model of a planar
nonlinear waveguide with the cubic nonlinearity, that features a
Bragg grating in the longitudinal direction, and is uniform along
the transverse coordinate. In the latter model, true solitons
cannot exist, as the spectrum of the system does not support a
full bandgap.

The objective of the present paper is to explore 2D and 3D
spatiotemporal solitons (which may be, strictly speaking,
``quasi-solitons", in the above sense, but feature completely
localized shapes) and their stability in the proposed medium. In
Section 2 we fix the mathematical form of the 2D version of the
model, analyze its spectrum, and apply the VA to the study of
solitons. In Section 3, direct numerical results demonstrating the
existence of very robust 2D solitons, and their delocalization
when the lattice strength $\varepsilon $ becomes too small, are
reported (the comparison with the VA prediction shows that the VA
provides for a crude approximation in the present model). In
Section 4, we additionally consider the effect of variation of the
nonlinearity coefficient along the propagation distance on 2D
solitons. Finally, numerical results for 3D solitons are collected
in Section 5, and the paper is concluded by Section 6.

\section{Formulation and variational analysis of the two-dimensional model}

The model of a planar waveguide corresponding to the outline given above is
based on the following variant of the 2D NLS equation for the local
amplitude $u(z,x,t)$ of the electromagnetic field:
\begin{equation}
iu_{z}-\frac{1}{2}u_{tt}+\frac{1}{2}u_{xx}+\varepsilon \cos (2x)u-|u|^{2}u=0,
\label{2dnlse}
\end{equation}
where $z$ is the propagation distance, $x$ is the transverse coordinate, and
$t$ is the reduced time, defined the same way as in fiber optics.
The signs in front of the GVD ($u_{tt}$) and cubic terms
correspond, as said above, to the normal GVD and self-defocusing
nonlinearity, $\varepsilon $ is the amplitude of the transverse
modulation of the refractive index (which is assumed sinusoidal,
but the results will be nearly the same for more realistic forms
of the modulation which correspond to the actual photonic-crystal
structure), and the period of the modulation is scaled to be $\pi
$. Dynamical invariants of Eq. (\ref{2dnlse}) are the norm of the
solution (in optics, it is the total energy),
\begin{equation}
N\equiv \int_{-\infty }^{+\infty }dx\int_{-\infty }^{+\infty
}dt|u(z,x,t)|^{2},  \label{N}
\end{equation}
together with the longitudinal momentum and Hamiltonian,
\begin{eqnarray}
P &=&i\int_{-\infty }^{+\infty }dx\int_{-\infty }^{+\infty }dt~u_{t}^{\ast
}u, \\
H &=&\int_{-\infty }^{+\infty }dx\int_{-\infty }^{+\infty
}dt\left[ \frac{1}{2}\left\vert u_{x}\right\vert
^{2}-\frac{1}{2}\left\vert u_{t}\right\vert
^{2}+\right.  \nonumber \\
&&\left. \frac{1}{2}|u|^{4}-\varepsilon \cos (2x)|u|^{2}\right] .
\end{eqnarray}

To find the linear spectrum of the model, one looks for solutions to the
linearized version of Eq. (\ref{2dnlse}) as
\begin{equation}
u(z,t,x)=\exp \left( ikz-i\omega t\right) F_{E}(x),  \label{linearized}
\end{equation}
where $k$ is a real propagation constant, $\omega $ is an arbitrary real
eigenvalue,\begin{equation} E\equiv \omega ^{2}-2k,  \label{E}
\end{equation}and $F_{E}(x)$ is a solution of the Mathieu equation,\begin{equation}
F^{\prime \prime }+2\varepsilon \cos \left( 2x\right) F+EF=0,
\label{Mathieu}
\end{equation}
corresponding to the eigenvalue $E$. It is commonly known that the Mathieu
equation gives rise to bandgaps in its own spectrum, i.e., to
forbidden intervals of the values of $E$, within which no regular
quasi-periodic solutions of Eq. (\ref{Mathieu}) can be found.
However, since all large values of $E$ ($E\gg \varepsilon $)
belong to the allowed band, where such solutions exist, it is
obvious that \emph{no} value of $k$ may fall in a forbidden
bandgap. Indeed, using Eq. (\ref{E}), one can construct \emph{any}
real value of the propagation constant as $k=\left( \omega
^{2}-E\right) /2$, taking very large $E$ and, accordingly, very
large $\omega $.

On the other hand, the same consideration suggests that, in some
cases, the necessary values of $E$ and $\omega $ may be very large
indeed. It was shown, in a general form, in Ref. \cite{Isaac} that
short-period waves corresponding to such large parameters, which
build up into a possible ``tail" attached to the soliton's ``body"
(that makes it a quasi-soliton), will have an exponentially small
amplitude, rendering the tail totally negligible (in particular,
it may be completely invisible in numerical solutions). Therefore,
it makes sense to look for ``practically existing" solitons in the
present model.

\begin{figure}[tbh]
\centerline{\includegraphics[width=8cm,height=6cm,clip]{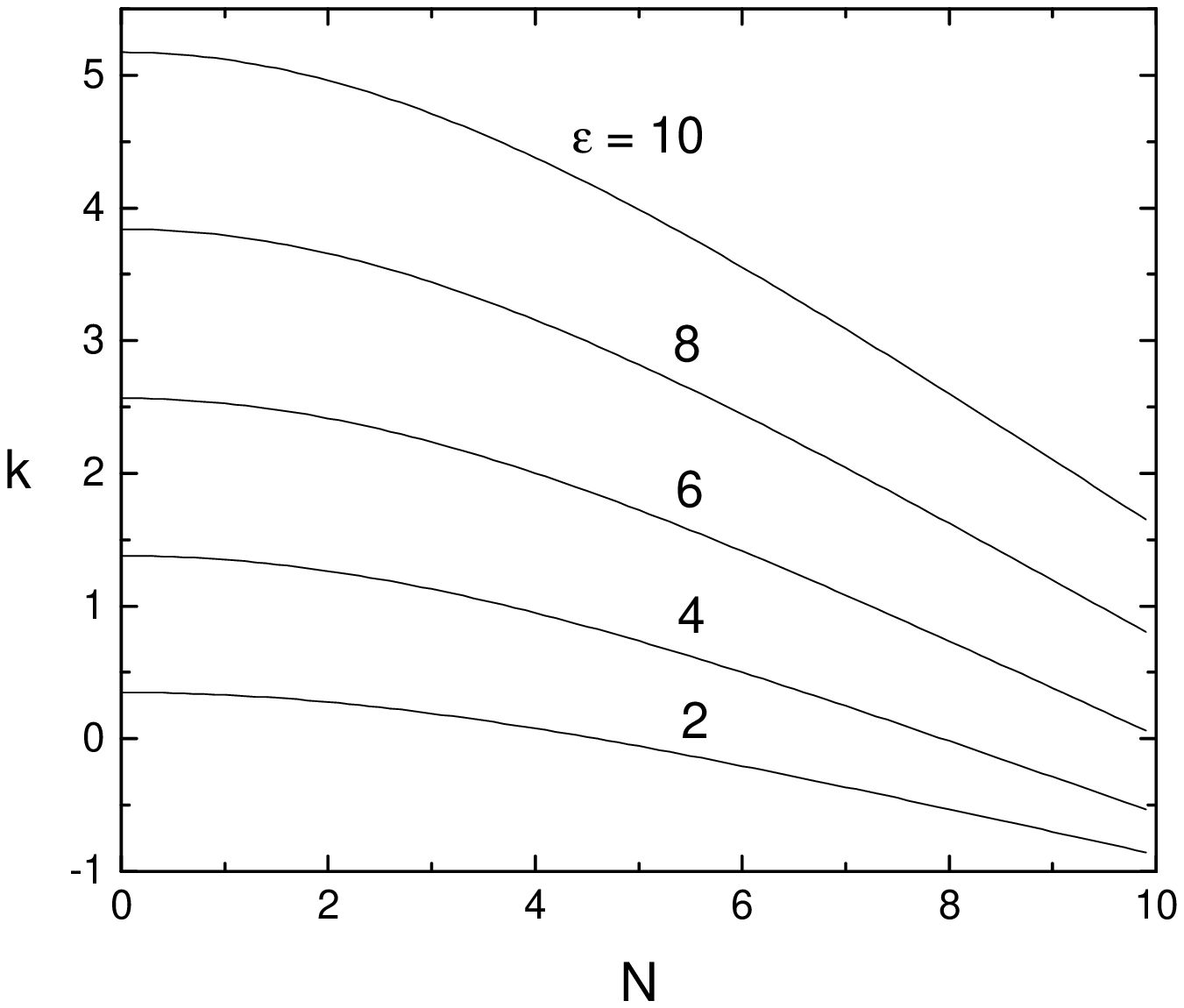}}
\caption{The propagation constant vs. norm for the 2D solitons, as predicted
by the variational approximation, Eq. (\protect\ref{2dva}), for different
values of the strength of the periodic potential $\protect\varepsilon $. The
negative slope, $dk/dN<0$, implies stability of the solitons according to
the Vakhitov-Kolokolov criterion.}
\label{fig1}
\end{figure}

We start searching for stationary solutions of the ``mixed" type, which, as
explained above, are expected to feature a gap-soliton (weakly localized)
shape along $x$ and strong ordinary localization in $t$, by adopting the
following variational ansatz,
\begin{equation}
u(z,x,t)=Ae^{ikz}\left[ x^{-1}\sin (ax)\right] \mathrm{sech}(\alpha t),
\label{2dansatz}
\end{equation}
where $a$ and $\alpha $ are, respectively, the transverse and longitudinal
inverse widths of the soliton, and $A$ is its amplitude. Using the obvious
Lagrangian representation of Eq. (\ref{2dnlse}) and well-known VA formalism
\cite{va}, one can readily derive the following equations for the parameters
of the ansatz,
\begin{equation}
a=\left( \frac{3\varepsilon }{1+(N/3\pi )^{2}}\right) ^{1/3},  \label{a}
\end{equation}
\begin{equation}
\alpha =\frac{aN}{3\pi },~A=\frac{N}{\sqrt{6}\pi },~k=\varepsilon
-\frac{\varepsilon ^{2/3}}{2\left( 27\pi ^{2}\right)
^{1/3}}\frac{27\pi ^{2}+5N^{2}}{\left( 9\pi ^{2}+5N^{2}\right)
^{2/3}},  \label{2dva}
\end{equation}
where $N$ is the norm defined by Eq. (\ref{N}). It follows from these
equations that the condition $dk/dN<0$, i.e., the necessary stability
condition, according to the above - mentioned VK criterion \cite{vk}, always
holds, as shown in Fig. \ref{fig1}.

The stability of the soliton families, predicted by the VK criterion as per
Fig. \ref{fig1}, is, generally, corroborated by direct numerical
simulations, see the next section (with a caveat that the VA predicts the
shape of the solitons in only a qualitatively correct form, as explained
below). However, it should also be mentioned that the applicability of the
VK criterion to gap solitons in lattice models has never been proven,
therefore one should apply this method with due care. In particular, it is
known that the gap solitons may be unstable in some cases when they are
expected to be VK-stable \cite{VK-}, which is not very surprising, as the VK
criterion ignores complex stability eigenvalues, that may give rise to
oscillatory instabilities. More perplexing is the fact that some gap-soliton
families in a lattice model with the cubic-quintic nonlinearity, that are
formally predicted to be VK-unstable, are in reality \emph{completely stable}
\cite{VK+}.

To complete the discussion of the VA, it is necessary to notice
that ansatz (\ref{2dansatz}) is irrelevant if it predicts $a\ll
1$, as it would mean that the soliton is very broad in the
$x$-direction, and it does not feel the underlying lattice
structure, $\cos (2x)$ in Eq. (\ref{2dnlse}). We therefore limit
the applicability of the VA by a (roughly defined) condition,
$a>1$. According to Eq. (\ref{a}), this implies that the potential
must be strong enough,
\begin{equation}
\varepsilon >\frac{1}{3}\left( 1+\frac{N^{2}}{9\pi ^{2}}\right) .
\label{crit}
\end{equation}

\section{Numerical results for two-dimensional solitons}

Direct simulations of Eq. (\ref{2dnlse}) (propagation in $z$) started with
an initial localized waveform, which we took as ansatz (\ref{2dansatz}) with
the parameters predicted by Eqs. (\ref{2dva}), or just a Gaussian with
rather arbitrary parameters -- for instance,
\begin{equation}
u(x,t,0)=A\exp [-\frac{a}{2}(x^{2}+t^{2})].  \label{Gauss}
\end{equation}
It was observed that the initial waveform undergoes intense evolution,
shedding off some radiation waves that were absorbed at edges of the
integration domain. The domain was large enough -- in most cases, $\left(
-8\pi ,+8\pi \right) $ in both directions ($x$ and $t$) -- so that the
solitons, which are typically well localized within a region $\left(
-5, +5 \right) $, see Figs. 2, 3, 7, and 9 below, are not affected by the
absorbers.
\begin{figure}[tbh]
\centerline{\includegraphics[width=8cm,height=6cm,clip]{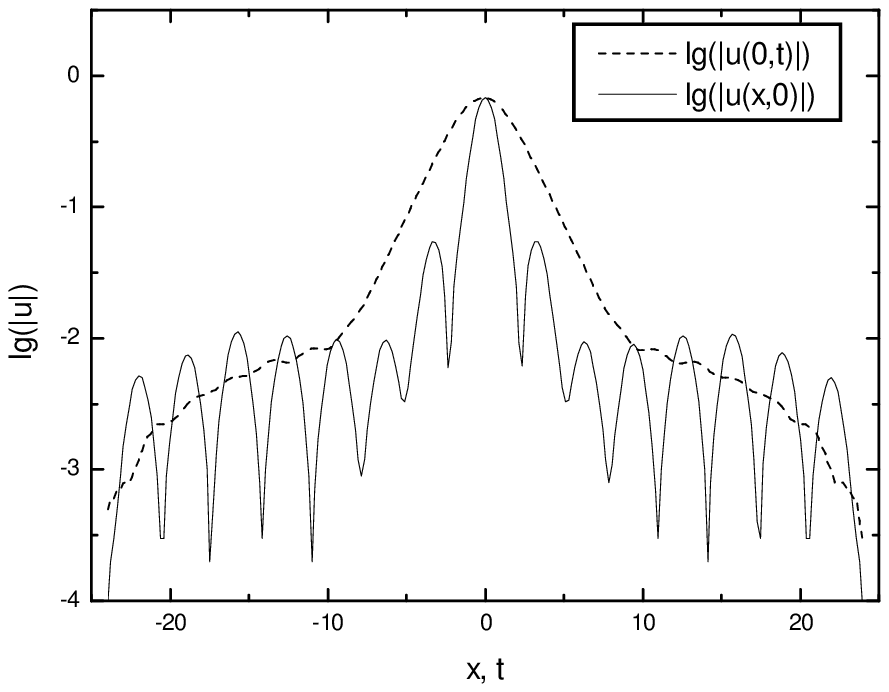}}
\caption{Cross sections of a typical established 2D soliton of the
semi-gap type. It has self-trapped from the initial configuration
(\protect\ref{Gauss}) with $A=1$ and $a=1$. The solid and dashed
lines represent, respectively, the sections along the transverse
directions $x$ at $t=0$, and temporal direction $t$ at $x=0$. The
shapes of the cross sections are displayed on the logarithmic
scale, to illustrate the fundamental observation of the practical
vanishing of the tails attached to the soliton's ``body".}
\label{fig2}
\end{figure}

\begin{figure}[tbh]
\centerline{\includegraphics[width=8cm,height=4cm,clip]{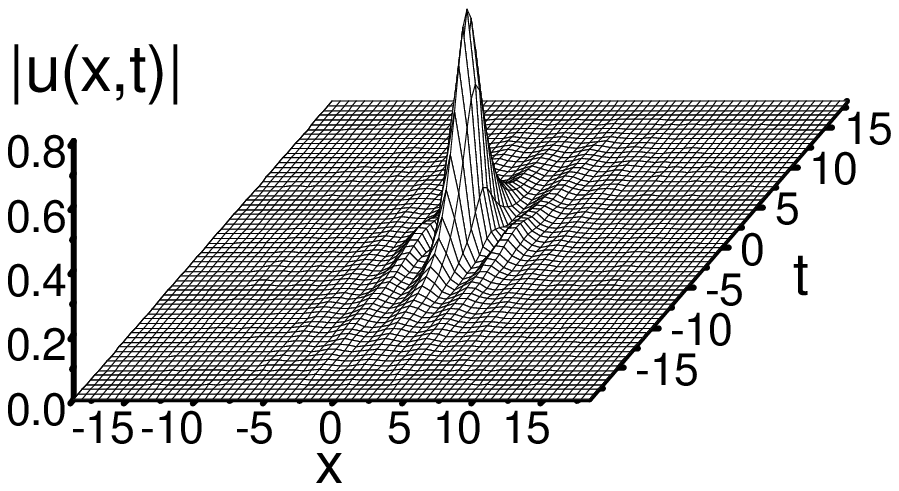}}.
\caption{A stable localized solution of Eq. (\protect\ref{2dnlse})
with $\protect\varepsilon =2.0$, produced from the initial
Gaussian pulse (\protect \ref{Gauss}) with the initial norm
$N_{0}=\protect\pi $. The amplitude, inverse width along the
temporal direction, and norm of this localized state are $A=0.74$,
$\protect\alpha =0.56$, and $N=2.29$, respectively. Note that
$27\%$ of the initial norm was lost with emitted radiation in the
course of the evolution. Notice that the soliton features no
extended tail.} \label{fig3}
\end{figure}

The evolution of the pulse ends with the establishment of a 2D
stationary soliton, \emph{without any visible tails}, as shown on
the logarithmic scale in Fig. 2. It is relevant to mention that,
as seen from Figs. \ref{fig2} and \ref{fig3}, the characteristic
sizes of the soliton in the $x$ and $t$ directions are on the same
order of magnitude, hence the values of the GVD coefficient and
its effective lattice-diffraction counterpart are, roughly, equal.

In fact, the solitons self-trap even from the initial
configurations that are quite different from their final shape,
which attests to strong robustness of the solitons. The soliton's
stability was then additionally tested against small random
perturbations, by simulating the evolution of the initial
configuration $U(x,t)=U_{0}(x,t)[1+\sigma u_{p}(x,t)]$, where
$U_{0}(x,t)$ is the numerically found soliton, $\sigma $ is a
small amplitude of the perturbation, and $u_{p}(x,t)$ is a random
function. An example of a stable localized state self-trapped from
the initial Gaussian (\ref{Gauss}) with $A=1$ and $a=1$, with the
norm $N_{0}=\pi $, is displayed in Fig. \ref{fig3}.

It should be said that direct comparison of the VA predictions
with the numerical results shows only a qualitative agreement: for
example, the VA predicts, with the same value of the norm, a
soliton whose width in the temporal direction exceeds the actual
width of the soliton in Fig. \ref{fig3}, by \ factor in excess of
$2$, and, accordingly, the amplitude of the numerically found
soliton significantly exceeds that predicted by the VA. Therefore,
the VA provides for only a crude approximation in this model;
nevertheless, its qualitative predictions, such as the stability
of the solitons predicted by the VK criterion, are correct. In
this connection, it is relevant to mention that no good version of
VA has been thus far proposed for gap solitons (this technical
issue was considered, in some detail, in Ref. \cite{Arik}), and
the solitons in the present model are still more complex objects.

\begin{figure}[tbh]
\centerline{\includegraphics[width=8cm,height=4cm,clip]{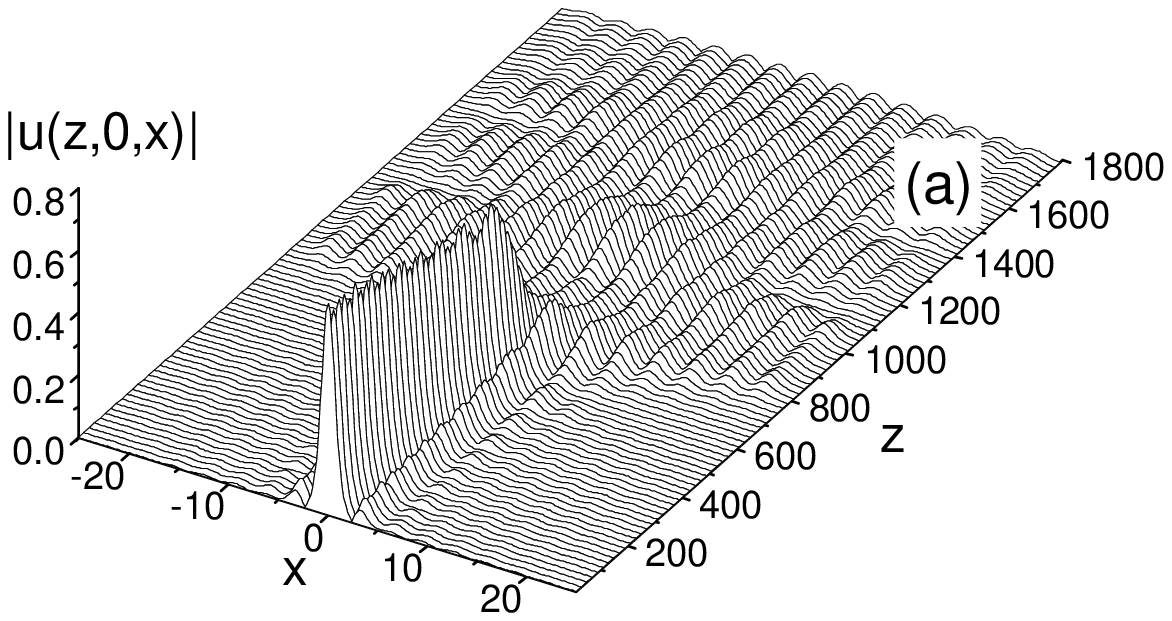}}
\centerline{\includegraphics[width=8cm,height=4cm,clip]{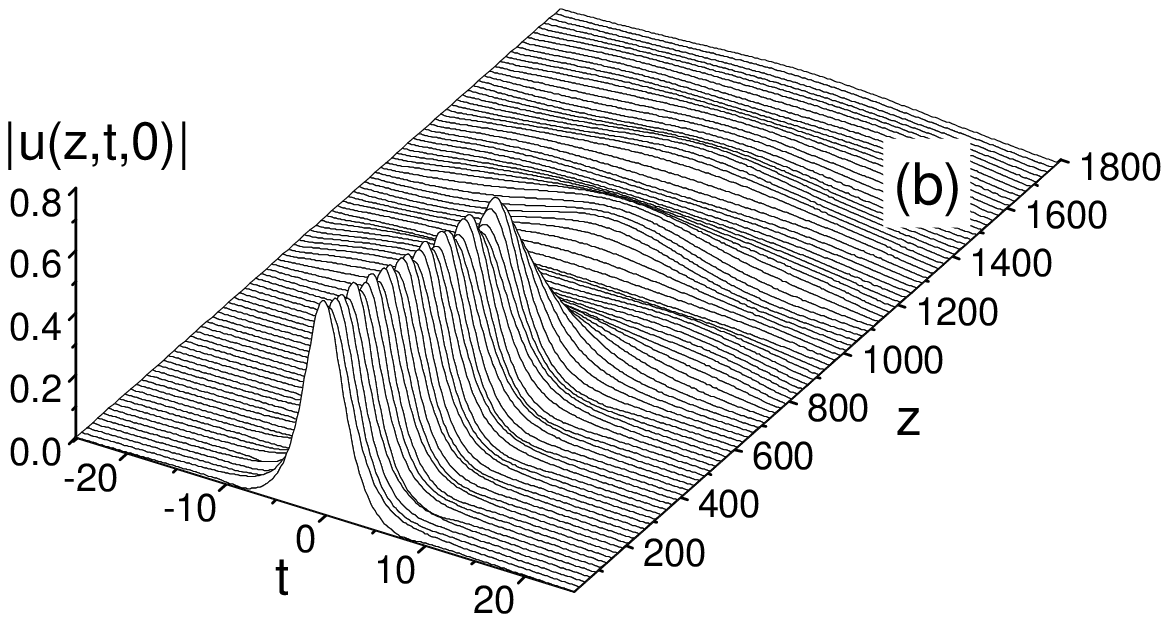}}
\caption{Dynamical disintegration of the localized state shown in
Fig. \protect\ref{fig3} as a result of a gradual decrease of the
strength of the periodic potential along the propagation distance,
so as $\protect\varepsilon (z)=\protect\varepsilon
_{0}(1-z/z_{\mathrm{end}})$, with $\protect\varepsilon _{0}=2$ and
$z_{\mathrm{end}}=1800$. Cross sections of the wave profile are
displayed: (a) along the transverse coordinate $x$; (b) in the
temporal direction $t$. The delocalization occurs around $z=650$,
at $\protect\varepsilon $ close to a critical value,
$\protect\varepsilon _{\mathrm{cr}}\simeq 1.3$.} \label{fig4}
\end{figure}

Further numerical analysis of the 2D model has demonstrated that,
following the known pattern of the delocalization transition of 2D
solitons in lattice potentials \cite{bs}, the soliton solutions
cease to exist when the strength of the periodic potential
$\varepsilon $ or the norm of the localized state $N$ fall below
some critical values. In this case, the localized waveform
undergoes disintegration, transforming into a quasi-linear
nonstationary extended state. This state keeps expanding until it
eventually hits edge absorbers, thus completely disappearing.
Naturally, the expansion occurs faster in the temporal (alias
longitudinal) direction, where it is not impeded by any potential
structure. Figure \ref{fig4} illustrates the disintegration of the
localized state (the one from Figure \ref{fig3}), following
gradual decrease of $\varepsilon $ along the propagation
direction. It is relevant to stress that the disintegration of the
soliton at small $\varepsilon $ is inevitable, as Eq.
(\ref{2dnlse}) with $\varepsilon =0$ has no 2D soliton solution,
unlike the Townes soliton, which would be a solution for the
equation with $\varepsilon =0$ and reverse sign in front of the
$u_{xx}$ term.

\begin{figure}[tbh]
\centerline{\includegraphics[width=8cm,height=6cm,clip]{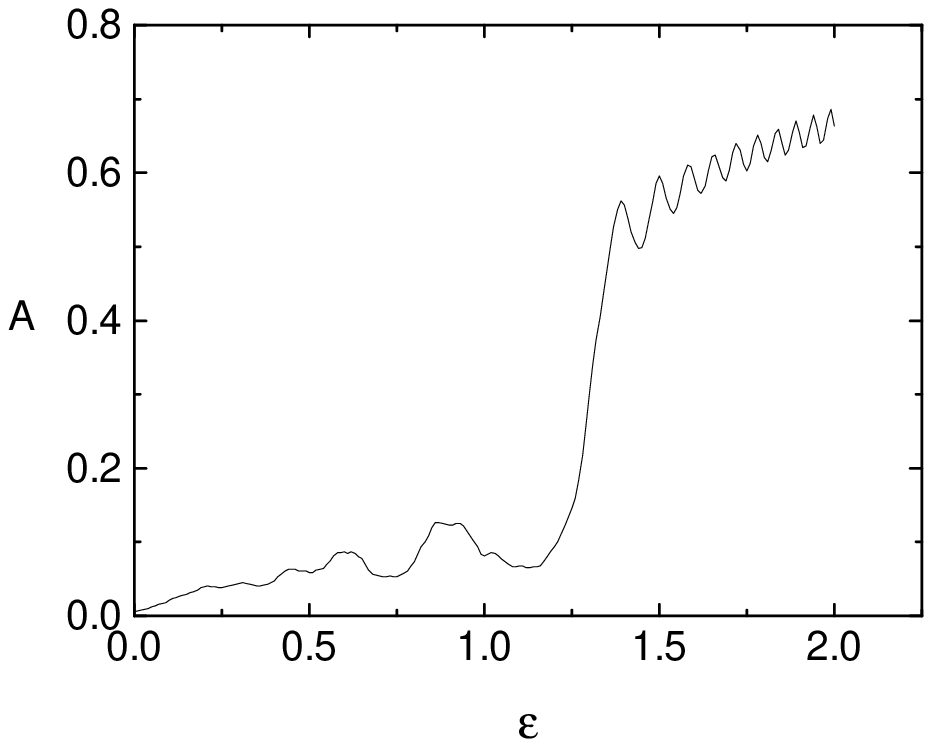}}
\caption{The amplitude of the 2D semi-gap soliton abruptly decays at the
point of the delocalization transition, which was displayed in Fig. \protect
\ref{fig4}. The present figure illustrates the steepness of the transition
at $\protect\varepsilon \simeq 1.3$, with the slowly decreasing strength of
the periodic potential, starting from $\protect\varepsilon =2.0$. }
\label{fig5}
\end{figure}

Detecting the delocalization transition of the semi-gap solitons at critical
values of the nonlinear coefficient (or rescaled norm) and/or the strength
of the periodic potential can be used to locate the lower border of their
existence region in the parameter space $\left( N,\varepsilon \right) $.
Actually, the transition from the localized state to the extended one is
quite steep (see Fig. \ref{fig5}), which allows quite accurate determination
of the critical values of the parameters. However, delineating the full
existence region of the semi-gap solitons is a harder problem, as the
limiting effect at large values of the norm, which determines the upper
border, is \emph{splitting} of the soliton (see below), rather than collapse
(singularity formation) in the case of lattice gap solitons with the
self-focusing nonlinearity \cite{bmsEPL}. Precise shapes of the existence
and stability domains of lattice gap solitons can be rather complex, as
shown in Ref. \cite{Kartashov} for the case of saturable nonlinearity.

Accurate determination of the full stability borders for the solitons in the
present model, going beyond the use of the VK criterion and collection of
typical examples of direct simulations, will be a subject of separate work.

Finally, Eq. (\ref{2dnlse}) features obvious Galilean invariance
in the longitudinal ($t$) direction, which makes it possible to
generate a \textit{boosted} soliton $u_{c}$, with an arbitrary
inverse-velocity shift $c$, from a given soliton $u$, as
\[
u_{c}(z,t)=u(z,t-cz)e^{-i\left( c^{2}/2\right) z-ict}.
\]
The use of such two pulses with the $c_{1}\neq c_{2}$ makes it possible to
study collisions between the moving solitons, which, however, should be a
subject of a separate work.

\section{Effects of nonlinearity modulation}

As the nonlinearity is a key factor necessary for the existence of solitons,
in this section we address response of the 2D solitons to variation of the
nonlinearity strength along the propagation coordinate, $z$. In nonlinear
optics, a variable nonlinearity coefficient can be created by dint of
different physical mechanisms, such as variation of a dopant concentration,
optically controlled photorefraction, or simply by using a variable
thickness of the planar waveguide.

\begin{figure}[tbh]
\centerline{\includegraphics[width=8cm,height=6cm,clip]{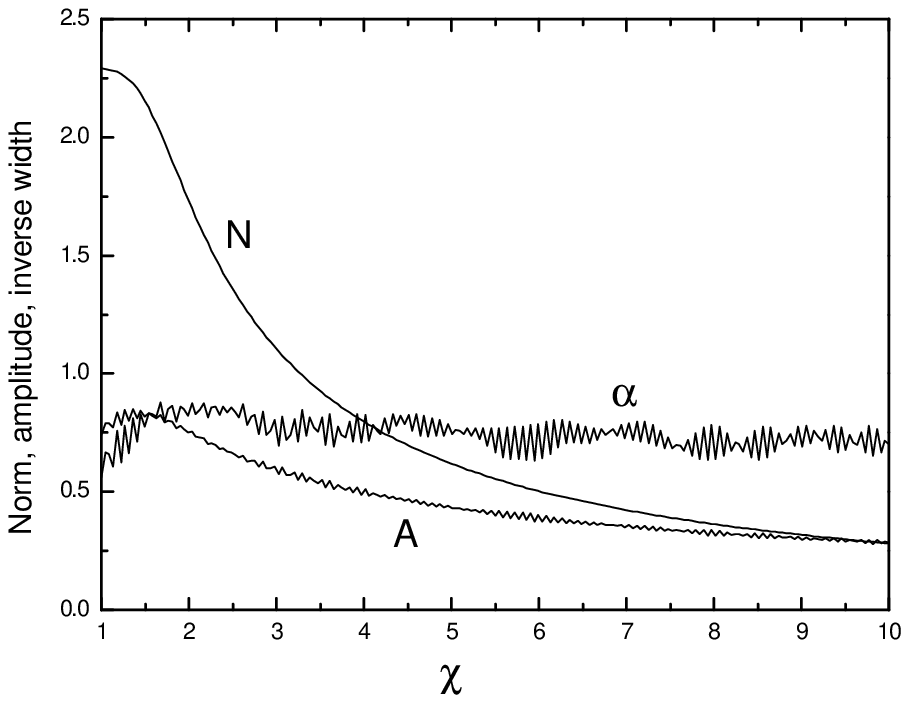}}
\caption{The evolution of the norm ($N$), amplitude ($A$), and
inverse temporal width ($\protect\alpha $) resulting from the slow
linear increase of the nonlinear coefficient $\protect\chi $ in
Eq. (\protect\ref{2dnlse}), $\protect\chi (z)=1+\protect\gamma
z/z_{\mathrm{end}}$,\textrm{\ }with $\protect\gamma =9$ and
$0<z<z_{\mathrm{end}}=2000$. Note significant decrease of the norm
$N$, opposed by little variation of $A$ and $\protect\alpha $. The
initial localized state is the same as in Fig.
\protect\ref{fig3}.} \label{fig6}
\end{figure}

Thus, we replace Eq. (\ref{2dnlse}) by its variant which includes
a variable nonlinear coefficient $\chi (z)$,
\begin{equation}
iu_{z}-\frac{1}{2}u_{tt}+\frac{1}{2}u_{xx}+\varepsilon \cos
(2x)u-\chi (z)|u|^{2}u=0.  \label{chi}
\end{equation}
Then, we follow the transformation of the soliton with slow increase or
decrease of $\chi (z)$. Figure \ref{fig6} displays the evolution of the
soliton's parameters as the nonlinearity coefficient gradually increases
ten-fold.

\begin{figure}[tbh]
\centerline{\includegraphics[width=8cm,height=4cm,clip]{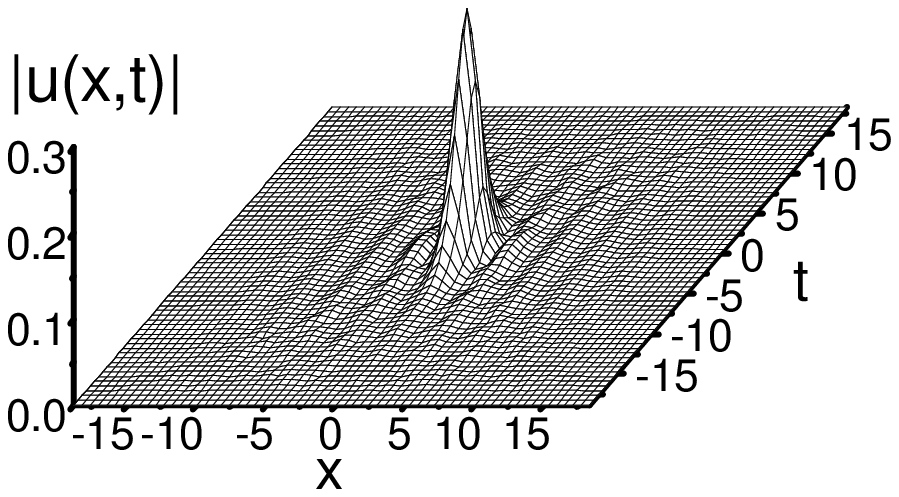}}
\caption{The final shape of the soliton from Fig. \protect\ref{fig3},
established after by slow increase of the nonlinearity by a factor of $10$,
as illustrated in the previous figure. Note that an extended tail attached
to the soliton does not appear in this case either.}
\label{fig7}
\end{figure}

In this case, Fig. \ref{fig7} shows that the final waveform,
corresponding to $\chi =10$, has not changed notably compared to
initial one (see Fig. \ref{fig3}), which implies that the increase
of the nonlinearity is countered by the loss of the norm. The
excess norm is shed off with linear waves which are absorbed on
the domain boundaries. It appears that the shape of the localized
wave, being weakly sensitive to the value of the norm (hence, to
the strength of the nonlinearity too), is actually fixed by the
strength $\varepsilon $ of the periodic potential. This
shape-invariance property of the mixed-type solitons is very
different from what is manifested by both ordinary solitons and
gap solitons per se, whose shapes are particularly sensitive to
the strength of the nonlinearity, at a fixed amplitude of the
periodic potential.

If the coefficient of nonlinearity slowly \emph{decreases}, as $\chi
(z)=1-z/z_{\mathrm{end}}$ with $z_{\mathrm{end}}=2000$, the disintegration
of the localized state is observed when falls to the level of $\chi \simeq
0.5$, resembling the picture in Fig. \ref{fig4}. Thus, we conclude that the
strengths of both the periodic potential and nonlinearity must exceed some
critical values in order to sustain the localized states. In this respect,
the multidimensional semi-gap solitons resemble regular gap solitons in
lattice potentials \cite{efremidis,bs}.

\begin{figure}[tbh]
\centerline{\includegraphics[width=8cm,height=6cm,clip]{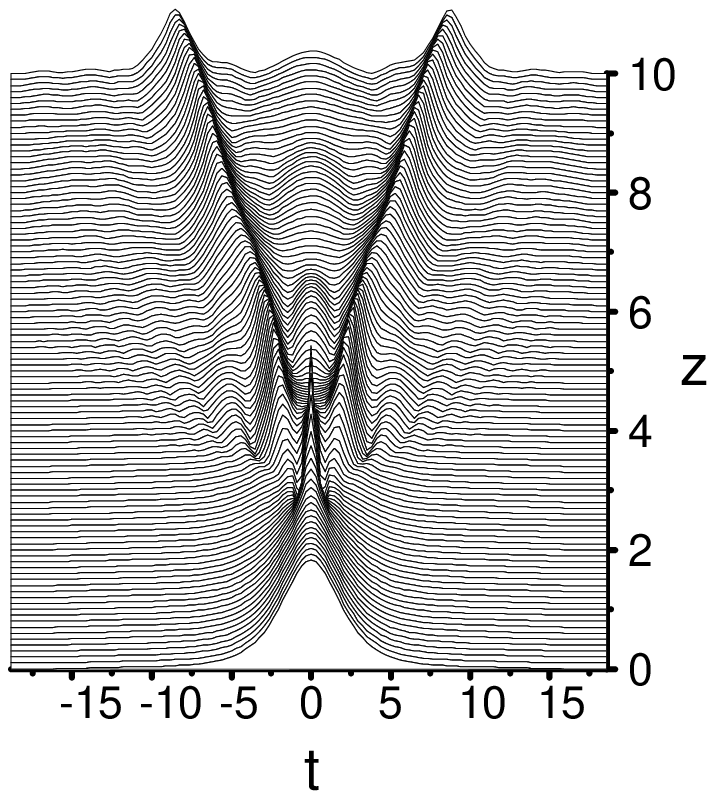}}
\caption{Splitting of the soliton from Fig. \protect\ref{fig2}
along the free direction $t$, following the rapid increase of the
nonlinear coefficient by a factor of $10$, which is performed by
setting $\protect\chi (z)=1+\protect\gamma \tanh
(4z/z_{\mathrm{end}})$ in Eq. (\protect\ref{chi}), with
$\protect\gamma =9$ and $z_{\mathrm{end}}=10$.} \label{fig8}
\end{figure}

One of characteristic features of ordinary solitons in 1D nonintegrable
systems is splitting of the soliton when the GVD coefficient \cite{grimshaw}
or nonlinearity \cite{Radik} is abruptly changed. In the present model, one
can observe a similar effect for 2D solitons. Figure \ref{fig8} displays an
example of the splitting of a soliton into two fragments, which then
separate along the free direction $t$, after the\ nonlinearity coefficient
was suddenly increased by an order of magnitude.

\section{The three-dimensional case}

\begin{figure}[tbh]
\centerline{\includegraphics[width=8cm,height=4cm,clip]{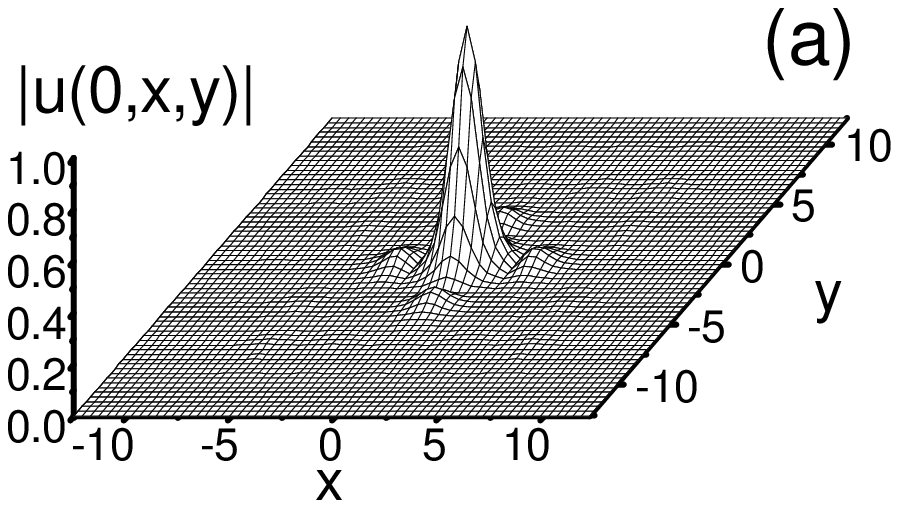}}
\centerline{\includegraphics[width=8cm,height=4cm,clip]{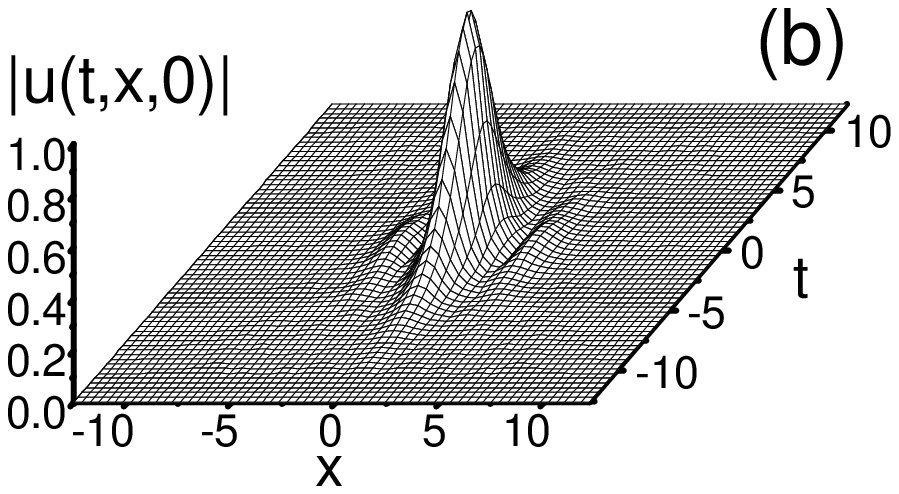}}
\caption{A three-dimensional stationary soliton solution to Eq.
(\protect\ref{3dnlse}) with $\protect\varepsilon =2.0$ is shown
through its two cross sections: (a) perpendicular to the free
direction, in the $t=0$ plane, and (b) parallel to the free
direction, in the $y=0$ plane. The soliton was obtained by direct
propagation in $z$ of an initial Gaussian with the norm
$N_{0}=2\protect\pi $. The norm of the established soliton is
$N=4.16$, i.e., a third of the initial norm was lost in the course
of the adjustment of the initial pulse to the stationary shape of
the 3D soliton.} \label{fig9}
\end{figure}

The 3D version of Eq. (\ref{2dnlse}) has a straightforward form,
\begin{equation}
iu_{z}-\frac{1}{2}u_{tt}+\frac{1}{2}(u_{xx}+u_{yy})+\varepsilon \lbrack \cos
(2x)+\cos (2y)]u-|u|^{2}u=0.  \label{3dnlse}
\end{equation}
Similarly to the 2D case, stationary solutions to Eq.
(\ref{3dnlse}) can be numerically found by using a Gaussian pulse
as the initial condition and propagating in $z$. Stability of the
3D solitons was verified by simulating the evolution of a soliton
with a random perturbation added to it. As well as in the 2D case,
the simulations were performed with the absorbers placed at
borders of the integration domain (under the condition that the
length of the domain was much larger than a characteristic size of
the soliton). As a result, it was concluded that robust 3D
solitons exist as generic solutions, without any visible tails
attached to them. An example of a stationary 3D soliton, which was
found to be quite robust in stability simulations, is displayed in
Fig. \ref{fig9}. In particular, Fig. \ref{fig9}(b) clearly
demonstrates that, as well as in the 2D case, the soliton has,
roughly, equal sizes in the spatial and temporal directions, i.e.,
the GVD and diffraction modified by the periodic potential play
equally important roles in supporting the solitons.

Besides the fundamental 3D pulses, such as the one displayed in
Fig. \ref{fig9}, the model can also support 3D solitons with
embedded \textit{vorticity}. Analogy with known results for
gap-soliton vortices in the 2D lattice models \cite{GapVortex}
suggests that the vortex solitons too may easily be stable in the
present model. However, detailed investigation of the vortex
solutions, as well as collecting systematic data about the family
of the fundamental 3D solitons, requires numerous runs of lengthy
simulations of the 3D equation, which is beyond the scope of the
present paper.

\section{Conclusion}

In this work, we have proposed a new type of the multidimensional
model in nonlinear optics. It combines self-defocusing
nonlinearity and normal group-velocity dispersion with periodic
modulation of the local refractive index in the one or two
transverse directions (in the 2D and 3D models, respectively).
Strictly speaking, multidimensional (spatiotemporal) solitons
cannot exist in media of this type, as the system's spectrum
contains no true bandgap. Nevertheless, solitons which seem as
completely localized ones are predicted by the variational
approximation, and found in direct simulations. These solitons are
solutions of a mixed type, as in the free (longitudinal, alias
temporal) direction they are regular solitons, while in the
transverse direction(s) they are objects of the gap-soliton type
(hence the solution as a whole was called a \textit{semi-gap
soliton}). The existence of the solitons requires that both the
norm of the solution (in other words, the nonlinearity strength
$\chi $) and the strength $\varepsilon $ of the spatially periodic
transverse potential exceed certain minimum values, otherwise the
pulses decay into linear waves. Actually, the solitons are much
more sensitive to $\varepsilon $ than to $\chi $.

The results reported in this paper call for further work, that should be
aimed at accurate identification of borders of the solitons' stability
regions, especially in the 3D model (which requires running very massive
simulations), investigation of collisions between solitons, that may move
freely in the longitudinal direction, and the study of vortex solitons in
the 3D case.

\section*{Acknowledgements}

B.B.B. thanks the Department of Physics at the University of
Salerno (Italy) for a two-year research grant. B.A.M. appreciates
hospitality of the same Department. The work of this author was
partially supported by the grant No. 8006/03 from the Israel
Science Foundation. M.S. acknowledges a partial financial support
from the MIUR, through the inter-university project PRIN-2003.

\end{document}